\documentclass[10pt,twocolumn]{article}
\usepackage{times,mathptm,graphicx,color}
\usepackage{anysize}
\marginsize{2cm}{2cm}{2cm}{3cm}

\begin{document}

\title{Is Quantum Search Practical?} % (v11b)}

\author{George F. Viamontes, Igor L. Markov, and John P. Hayes \\
  \{gviamont, imarkov, jhayes\}@eecs.umich.edu \\
  The University of Michigan, Advanced Computer Architecture Laboratory \\
  Ann Arbor, MI~~48109-2122, USA
}
\date{ }
\vspace{-2mm}
\maketitle
\thispagestyle{empty}

\abstract{
\vspace{-2mm}
%\addtolength{\baselineskip}{2mm}
  Quantum algorithms and circuits can, in principle, outperform the best
non-quantum (classical) techniques for some hard computational
problems. However, this does not necessarily lead to useful
applications. To gauge the practical significance of a quantum
algorithm, one must weigh it against the best conventional
techniques applied to useful instances of the same problem.
Grover's quantum search algorithm is one of the most widely
studied.
   We identify requirements for Grover's algorithm to be useful in practice:
    (1) a search application S where classical methods do
not provide sufficient scalability; (2) an instantiation of Grover's algorithm
Q(S) for S that has a smaller asymptotic worst-case runtime than any classical
algorithm C(S) for S; (3) Q(S) with smaller actual runtime for practical
instances of S than that of any C(S).
   We show that several commonly-suggested applications fail to satisfy
these requirements, and outline directions for future work on
quantum search.

\vspace{-2mm}
\section{Introduction}
\vspace{-2mm}

 There is growing interest within the electronic design automation (EDA)
 community in quantum mechanics.  This interest
 is mostly motivated by the fact that experimental transistors have
 already reached the scale of several atoms, where quantum-mechanical
 effects are not only well-pronounced, but can perform useful
 functions. A radical new approach to harnessing these effects suggests storing
 information in quantum states and manipulating it using quantum-mechanical
 operators \cite{NielsenC}. Such quantum states may be carried by
 electron or nuclear spins in molecules, ions trapped in magnetic fields,
 polarizations of photons or quantized currents in superconductors.
 Quantum-mechanical operations can be performed by RF pulses, optically-active media
 and single-photon detectors, or Josephson junctions. Quantum states are often measured
 by single-photon detectors, e.g., by exciting an ion and forcing it to emit a photon.
 Surprisingly, some algorithms in terms of quantum states have
 better worst-case asymptotic complexity than best known
 conventional algorithms.

 The basic unit of quantum information is the {\em qubit} (quantum bit),
 conventionally written in the form ${|\psi> = c_0|0> + c_1|1>}$,
 where the coefficients $c_0$ and
 $c_1$ are complex numbers related to the probability of the qubit
 being 0 or 1. This expression for $|\psi>$ is interpreted as a
 superposition state that is both 0 and 1 simultaneously.
 A $k$-qubit system can be modeled by a complex-valued vector of the form
 $|q_1,q_2,...,q_k>$ which denotes a superposition of $2^k$
 non-quantum or "classical" states, and implies the presence of
 a kind of massive parallelism in
 the quantum state.  This parallelism is a key source of
 speedup  in quantum computing, but it also
 makes simulation of such computations on conventional
 computers exceedingly difficult.
 In general, mathematical modeling techniques for quantum effects entail the
 use of very large complex-valued matrices for basic (gate) operations
 and either state-vectors or density-matrices for the quantum states.
 In a way, complex-valued vectors generalize bit-strings, and matrices
 generalize truth tables.

 While based on very different physics, quantum circuits to some
 extent resemble classical logic circuits, at least in the sense
 that they can be drawn using circuit diagrams. A particularly successful
 application of quantum information processing is found in cryptography,
 and relies on a postulate from quantum mechanics asserting that a quantum
 state is destroyed when it is measured. With additional
 arrangements for quantum communication (typically performed by sending
 single photons through fiber optic cable), one can guarantee that
 any attempt to eavesdrop is fruitless and also detected.
 Operational quantum cryptography systems are commercially available
 from MagiQ Technologies in the U.S. and IdQuantique in Europe.

    Insights by Feynman, Deutsch, Shor and others suggest that massive
  speed-ups in computing can be achieved by exploiting
  quantum-mechanical effects such as superposition (quantum
  parallelism) and interference \cite{NielsenC}.
 A quantum algorithm typically consists of applying quantum gates
 to quantum states, but since the input to the algorithm may be non-quantum,
 i.e., normal classical bits,
  it only affects the selection of quantum gates. After all gates
  are applied, quantum measurement is performed and produces the
  non-quantum output of the algorithm.  Deutsch's algorithm,
  for instance, solves a certain artificial problem in fewer steps
  than any classical (non-quantum) algorithm can, and its relative
  speed-up grows with the problem size. % \cite{Deutsch}.
  However, Bennett et al. have shown in \cite{BenettBBV97} that quantum algorithms
  are unlikely to solve NP-complete problems in polynomial time,
  although more modest speed-ups remain possible.
  Shor designed a fast (polynomial-time) quantum algorithm for number
  factoring --- a key problem in cryptography that is not believed to be NP-complete. % \cite{NielsenC}.
  No classical polynomial-time algorithm for number factoring is known,
  and the problem seems so hard that the security of the RSA code used
  on the Internet relies on its difficulty. If a large and error-tolerant
  quantum computer were available today, running Shor's algorithm on it
  could compromise e-commerce.

  Another important quantum algorithm due to Grover
  \cite{Grover97,NielsenC,Boyer96} searches an unstructured
  ``database'' to find $M$ records that satisfy a given criterion.\footnote{
  A brief description of Grover's algorithm and its implementation
  is given in the Appendix.}
  For any $N$-element database it takes $\sim\sqrt{N/M}$ evaluations of
  the search criterion  (queries to an oracle) on database elements,
  while classical algorithms provably need at least $\sim N$ evaluations for some inputs.
  Despite the promise of the theory, it is by no means clear whether, or how
soon, quantum-computing methods will offer better performance in
useful applications \cite{Preskill98}. As explained later,
traditional complexity analysis of Grover's algorithm
\cite{Grover97,NielsenC,Boyer96} does not consider the complexity
of oracle queries. The query process is simply treated as a
``black-box'', thus making Grover's algorithm appealing because it
needs fewer queries than classical search. However, with a
sufficiently time-consuming query process, Grover's algorithm can
become nearly as slow as a simple (exhaustive) classical search.

  Grover's algorithm must also compete with advanced classical search techniques
  in applications that use parallel processing \cite{Zalka00}
  or exploit problem structure, often implicitly. In this work, we identify
  and analyze several  requirements necessary for Grover's search to be useful in
  practice:

  \begin{enumerate}
 %\vspace{-2mm}
   \item A search application $S$ where classical methods do not
         provide sufficient scalability.
% \vspace{-2mm}
   \item An instantiation $Q(S)$ of Grover's search for $S$
         with an asymptotic worst-case runtime which is less
         than that of
         any classical algorithm $C(S)$ for $S$.
% \vspace{-2mm}
   \item  A $Q(S)$ with an actual runtime for practical instances of $S$,
          which is less than that of any $C(S)$.%
%vspace{-2mm}
  \end{enumerate}

     We argue that real-life database applications rarely satisfy Requirement 1.
     Requirement 2 limits the runtime of quantum-oracle queries,
     and raises subtle hardware design issues.
%     Additionally, recent progress in classical algorithm theory for several
%     key search problems creates a strong competition for quantum search.
     Requirement 3 points to approximation algorithms and heuristics for $C(S)$,
     as well as fast, adaptive simulation of $Q(S)$ on classical computers.
     We describe a simulation methodology for evaluating potential
     speed-ups of quantum computation for specific instances of $S$.
     We demonstrate that search problems often contain a great deal of
     structure, whose possible use by classical algorithms must be
     factored into the evaluation of Requirements 2 and 3. This
     analysis suggests several directions for future research.

Section \ref{sec:background} reviews quantum search methods, and
Section \ref{sec:apps} discusses the scalability issue mentioned
in Requirement 1 for major applications. In Section
\ref{sec:query_complex}, we study the runtime of Grover's search
in the context of Requirement 2. Section  \ref{sec:simulation}
demonstrates that efficient classical simulation of quantum search
can have runtime which is competitive with Grover's algorithm in
useful instances. Section \ref{sec:showdown} provides a number of
comparisons to classical algorithms that meet Requirements 2 and
3. Conclusions and directions for future work on quantum search
are summarized in Section \ref{sec:conclusions}.

%\vspace{-2mm}
\section{Quantum Search}
%\vspace{-3mm}
\label{sec:background}

%  Below we summarize results due to Grover \cite{Grover97} including
%  his quantum algorithm and a comparison against best classical algorithms.
%  We also outline recent work on simulation of quantum circuits
% \cite{ViamontesMH03}.

  To search a (large) database used in some particular application $S$,
  Grover's algorithm must be supplied with two different
  kinds of inputs that depend on $S$: (i) the database itself, including its
  read-access mechanism; and (ii)  the search criteria, each of which is
  specified by a black-box predicate or oracle $p(x)$ that can be
  evaluated on any record $x$ of the database. The algorithm then
  looks for an $x$ such that $p(x)=1$. In this
  context, $x$ can be addressed by a $k$-bit string, and the database
  can contain up to $N=2^k$ records.

  Classically, we may evaluate or  query  $p(\cdot)$ on one input at a
  time.  In the quantum domain, however, if
  $p(\cdot)$ can be evaluated on either $x$ or $y$, then it can also
  be evaluated on the superposition $(x+y)/\sqrt{2}$, with the result
  $(p(x)+p(y))/\sqrt{2}$. This quantum parallelism
  enables search with $\sqrt{N}$ queries \cite{Grover97}. If $M$
  elements satisfy the predicate, then $\sqrt{N/M}$ queries suffice
  \cite{Boyer96}. Note that the parallel evaluation of $p(\cdot)$
  requires a superposition of multiple bit-strings at the input, which
  can be achieved by starting in the $|00\ldots 0\rangle$ state and
  applying the Hadamard gate $H$ on every qubit. This, of course,
  requires that $p(\cdot)$ can interpret a bit-string as an index of a
  database record.
% Assuming that $p(\cdot)$ reads every input qubit,
%  each query must take at least linear time.

   Several variants of Grover's algorithm are known, including
  those based on quantum circuits and different forms of adiabatic
  evolution. However, as discussed in \cite{RolandC03},
  all are closely related to the original algorithm
  and have similar computational behavior.

    For comparison, consider a classical deterministic algorithm
  for unstructured search, assuming that parallel processing techniques
  are not employed \cite{Zalka00}.  It requires making many queries,
  because an unsuccessful evaluation of $p(x)$ does not, in general,
  yield new information about records other than $x$. Therefore,
  one may need anywhere from 1 to $N$ queries, depending on the input ---
  $N/2$ on average. We must make queries until one of them is successful,
  and we cannot take advantage of an unsuccessful query.  Thus, every
  deterministic algorithm must visit the database records one by one, in
  some order, and independently try up to $2^k$ database records until
  a desired record is found. Randomized algorithms can pick records at
  random, and have an edge over deterministic algorithms when many
  records satisfy $p(\cdot)$. That is because for any input
  approximately $\frac{N}{2M}$ queries suffice with very high
  probability.  However, this improvement is not comparable with the
  quadratic speedup offered by Grover's algorithm.

\vspace{-2mm}
\section{Application Scalability}
\vspace{-2mm} \label{sec:apps}

  While Grover's algorithm relies on quantum mechanics,
  it nevertheless solves a classical search problem and
  competes with advanced classical search techniques
  in existing as well as new applications. Existing applications
  include Web search engines, very large databases for
  real-time processing of credit-card transactions,
%   off-line search for patterns of identity fraud,
  analysis of high-volume astronomic observations, etc.
  Such databases explicitly store numerous pieces
  of classical information (records).  Another class of existing
  applications is illustrated by code-breaking and Boolean satisfiability,
  where the input is a mathematical function $p(x)$, specified concisely
  by a formula, algorithm or logic circuit.
  One seeks the bits of $x'$ such that $p(x')=1$,
  which may represent a correct password or encryption key.
  The database of all possible values of $x$
  is {\em implicit} and does not require large amounts of memory.

  {\em Explicit  databases} in existing applications are often too large
  to fit in the memory of one computer. They are distributed through the network
  and searched in parallel. Records can be quickly added, copied and
  modified. Distributed storage also facilitates redundancy, back-up
  and crash recovery.  The records in such databases correspond to
  physical objects (sensors, people or Web pages), and this tends to
  limit typical growth rates of databases. Explicit databases may
  {\em temporarily} experience exponential growth, as exemplified by the World
  Wide Web, and yet existing search infrastructures appear scalable
  enough for such applications, as illustrated by the continuing success of
  the Web search engine {\tt google.com}.
%  even through a period when the Web exhibits exponential growth.
  Grover's algorithm, on the other hand, is not well suited
  to searching explicit databases of the foregoing kind because it demands
  a quantum superposition of all database records.
  Creating such a superposition, or using a superposition of indices in that
  capacity, seems to require localizing classical records in one place,
  which is impractical for the largest explicit databases.
  % To this end, popular descriptions
  % of Grover's algorithm that involve searching a telephone book do not necessarily
  % lead to useful practical applications.
  % Circumventing these problems is an interesting but difficult line of research.

  Consequently, Grover's search algorithm seems confined to {\em implicit
  databases}, where it also faces serious competition from
  classical parallel methods \cite{Zalka00}. This application
  class includes cryptographic problems, which are amenable to
  classical massively-parallel computation. For instance,
  the DES Challenge II decryption problem has been solved
  in one day by a custom set of parallel processors
  built by the the Electronic Frontier Foundation and {\tt distributed.net}
  for $\$250,000$.
  %% For example, Xilinx supplies general-purpose
%%   field-programmable gate array (FPGA) chips containing over a million
%%   gates \cite{Xilinx03}. We also note that processor caches, such as
%%   those in Intel's Pentium chips, are essentially hardware-optimized
%%   search engines --- they reduce a processor's accesses to main memory
%%   by storing and looking up frequently used data. Such custom-designed
%%   circuits are both well-studied and orders-of-magnitude faster than
%%   FPGAs (but much more expensive).
%  Large numbers of parallel {\em quantum} processors
%  may be very difficult to connect and synchronize.
%  However, Grover and Rudolph \cite{GroverR} recently showed that,
%  in theory, such form of parallelism makes some quantum algorithms
%  no better than variants of Grover's search.
%%Note, however, that such specific cryptography challenges
%%seek to estimate progress in hardware and software,
%%and do not capture utterly hopeless tasks.

  Implicit search applications typically exhibit
  exponential scalability, e.g., adding an extra bit to an
  encryption key doubles the key
  space. This cannot be matched in principle by the linear scalability
  of classical parallel processing techniques (i.e., adding hardware).
  Therefore, we believe that these applications meet Requirement 1,
  and thus are potential candidates for practical
  quantum search tasks.

\vspace{-2mm}
\section{Oracle Implementation}
\vspace{-2mm} \label{sec:query_complex}

Although the oracle function $p(\cdot)$ in Grover's algorithm can
be evaluated on multiple inputs simultaneously, the description of
$p(\cdot)$ is usually left unspecified
\cite{Grover97,NielsenC,Boyer96}. To actually implement Grover's
algorithm for a particular search problem, one must explicitly
construct $p(\cdot)$. Several pitfalls are associated with this
important step, and are related to the complexity of $p(\cdot)$.

The first problem is that to query $p(\cdot)$ using quantum
parallelism, one must implement $p(\cdot)$ in quantum hardware.
This hardware can take a variety of different logical and physical
forms \cite{NielsenC}. If a quantum implementation of $p(\cdot)$
is derived from classical hardware design techniques, the circuit
size of the classical and quantum implementations may be similar.
Circuit size is estimated by the number of logic operations
(gates) employed, and computation time by the maximum depth of the circuit.
However, if these numbers for an $N$-item database scale much worse than
$\sqrt{N}$, then both classical and quantum searching will be
dominated by the evaluation of $p(\cdot)$, diminishing the
relative value of the quantum speed-up on the log-scale.

A more subtle problem is the complexity of {\em designing}
hardware implementations of $p(\cdot)$. Even if a given $p(\cdot)$
can theoretically be implemented without undermining the relative
speed-up of Grover's algorithm, there may be no practical way to
find compact classical or quantum implementations in a reasonable
amount of time. In classical electronic design automation (EDA),
finding small logic circuits is an enormously difficult
computational and engineering task %\cite{HachtelS}
that requires synergies between circuit designers and expensive
design software. Automatic synthesis of small quantum circuits
appears considerably harder as some formulations allow gates whose
function depends on continuous parameters \cite{ShendeMB04},
rendering discrete methods irrelevant.

While Requirement 2 is satisfied by Grover's algorithm in
principle, satisfying it by a {\em significant margin} on the
log-scale may be difficult in many cases because a small-circuit
implementation of the oracle-function $p(\cdot)$ may not exist, or
may require an unreasonable effort to find.

%% Another issue that arises in Grover's algorithm is the
%%   problem of reading in an arbitrary predicate $p(\cdot)$. Since the
%%   predicate is assumed to be a black-box, the structure of $p(\cdot)$
%%   may be unknown, making it difficult for an implementation of the
%%   algorithm to take a $p(\cdot)$ as an input. If, on the other hand, a
%%   description of $p(\cdot)$ is available, lower bounds for classical
%%   search using the same predicate are invalidated because faster,
%%   specialized algorithms may exist which exploit the predicate's
%%   structure (see the next section for examples). This predicate input
%%   dilemma may be circumvented in a communications setting where $x$
%%   is sent to another party, who performs the query $p(x)$ and sends
%%   back the result without revealing any additional information about
%%   the predicate. However, this scenario merely shifts
%%   the burden of having a predicate description to the second party.

\begin{figure}[tb]
\begin{center}
\includegraphics[width=6cm]{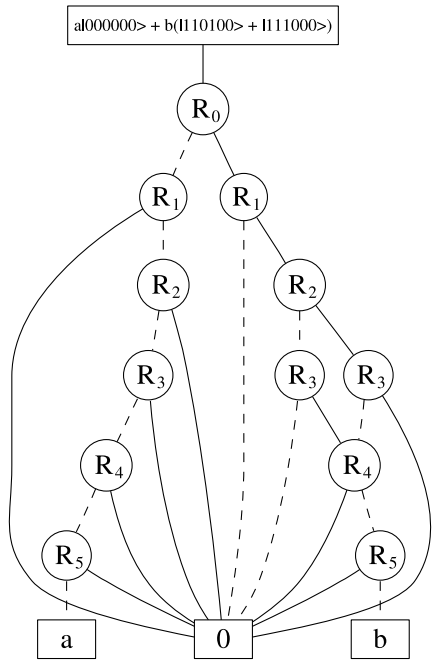}
\end{center}
%\vspace{-5mm}
 \caption{\label{fig:quidd}
%\small
%\addtolength{\baselineskip}{2mm}
  A 5-qubit
state-vector in the QuIDD data structure.  Each decision variable
$R_i$ corresponds to a bit $i$ in the binary encoding of
  indices in the vector. Dashed lines model $0$s assigned to the
  index bit, and solid lines model $1$s.  Top-down paths represent
  the 32 entries of the state-vector; a path may capture multiple
  entries with equal values.  }
%\vspace{-5mm}
\end{figure}

\vspace{-3mm}
\section{Classical Simulation}
\vspace{-3mm} \label{sec:simulation}

As in the case of Grover's algorithm, quantum computation is often
represented in the quantum-circuit formalism, which is described
mathematically using linear algebra \cite{NielsenC}. Qubits are
the fundamental units of information; $k$-qubit quantum states can
be represented by $2^k$-dimensional vectors, and gates by square matrices
of various sizes. The parallel composition of gates corresponds to
the tensor (Kronecker) product, and serial composition
to the ordinary matrix product.
Like classical circuits, quantum circuits can be conveniently
simulated by computer software for analysis or design purposes.
A quantum circuit can be simulated naively by a sequence of
$2^k\times 2^k$-matrices that are applied sequentially to a state
vector.  This reduces quantum simulation to standard linear
algebraic operations with exponentially sized matrices and
vectors.  Measurement is simulated similarly. Since Grover's
algorithm only requires $k$ qubits for a database of $N=2^k$
records, it is clear that a naive classical simulation of Grover's
algorithm would be exponentially worse than an actual quantum
circuit that implemented the algorithm.

  The linear-algebraic formalism does not differentiate between
  structured and unstructured data. However, the state vectors and
  gate matrices that appear in typical quantum simulations are anything
  but unstructured. In particular, they compress very well when simulated
  using the QuIDD data structure \cite{ViamontesMH03}. A QuIDD is a directed
  acyclic graph with one source and multiple sinks, where each sink is
  labeled with a complex number. Matrix and vector elements are
  modeled by directed paths in the graph, as illustrated in Figure
  \ref{fig:quidd}. Linear-algebraic operations can then be implemented
  by graph algorithms in terms of compressed data representations.
  The use of data-compression may substantially reduce simulation runtime
  for specific applications, especially those dealing with non-random data
  and circuits. This suggests a {\em test-by-simulation} approach to
  identify violations of Requirement 3. Indeed, polynomial-time
  simulation techniques were proposed for circuits with restricted
  gate types \cite{Gottesman98,Valiant02} and for ``slightly entangled''
  quantum computation \cite{Vidal03}. However, these results have not
  been applied to quantum search.

  We have found that QuIDDs enable a useful class of quantum circuits to be
  simulated using time and memory that scale polynomially with the number of
  qubits \cite{ViamontesMH03}. All the components of Grover's algorithm,
  except for the application-dependent oracle, fall into this class.
  In fact, we have also proven that a QuIDD-based simulation of
  Grover's algorithm requires time and memory resources that are
  polynomial in the size of the oracle $p(\cdot)$ function represented
  as a QuIDD \cite{ViamontesMH03}. Thus, if a particular $p(\cdot)$
  for some search problem can be represented as a QuIDD using
  polynomial time and memory resources (including conversion of an original
  specification into a QuIDD), then classical simulation of
  Grover's algorithm performs the search nearly as fast
  as an ideal quantum circuit.
  If a practical implementation of an oracle function $p(\cdot)$
  is known, it is straightforward to represent it in QuIDD form,
  since all relevant gate operations are defined for QuIDDs.
  Once $p(\cdot)$ is captured by a QuIDD, a QuIDD-based
  simulation requires $\sqrt{N/M}$ queries, just like an actual
  quantum computer \cite{ViamontesMH03}. If the size of the QuIDD for
  $p(\cdot)$ scales polynomially for $k$-qubit instances of the search problem,
  then Grover's algorithm offers no speed-up for the given search problem.

  We implemented a generic QuIDD-based simulator called QuIDDPro
  in the C++ programming language \cite{ViamontesMH03}. This
  simulator can be used in practice to perform the test by simulation
  just described. Figure \ref{fig:runtime} presents runtime results
  for QuIDDPro simulating Grover's algorithm for the search problem
  considered in the seminal paper by Grover \cite{Grover97}. The
  oracle function for this search problem returns $p(x)=1$ for one
  item in the database. In all such cases, the QuIDD for $p(\cdot)$ has
  only $k$ nodes, and QuIDD-based simulation is empirically as fast as
  an actual quantum computer. Memory usage is only a few megabytes
  and grows linearly with the number of qubits \cite{ViamontesMH03}.
  Hence this particular search problem {\em fails} to meet
  Requirement 3, and so does not benefit by being
  implemented on a quantum computer.

\begin{figure}[tb]
\begin{center}
\includegraphics[width=8cm]{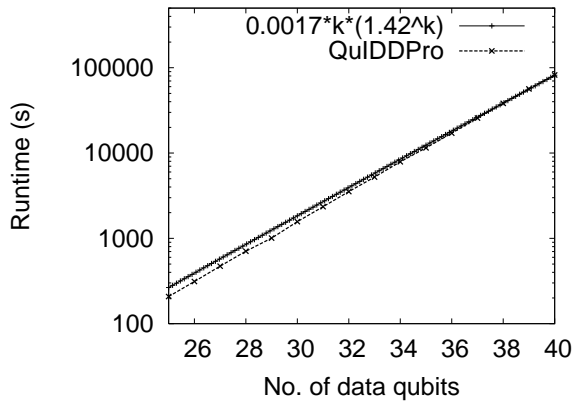}
\end{center}
\vspace{-4mm}
 \caption{\label{fig:runtime} %\small
%\addtolength{\baselineskip}{2mm}
   For $k$ qubits, an actual quantum computer implementing Grover's
   algorithm must perform at least $ck(\sqrt{2})^k$ steps where $c$ is a
   constant. For the type of predicates originally considered by Grover
   \cite{Grover97},
   empirical runtime of QuIDDPro simulation fits very well to this formula
   (plotted on a log-log scale), and the value of $c$ is small.}
\vspace{-1mm}
\end{figure}

  %% The surprising efficiency of this simulation
%% technique calls into question the value of implementing Grover's
%% algorithm on a quantum computer.  All the quantum gates used in
%% Grover's algorithm, including the Hadamard gate, can be
%% represented and manipulated using time and memory that is at most
%% linear in the number of qubits \cite{ViamontesMH03}. Once
%% $p(\cdot)$ is captured by a QuIDD, a QuIDD-based simulation
%% requires $\sqrt{N/M}$ queries, just like an actual quantum
%% computer \cite{ViamontesMH03}. The runtime of simulation
%% additionally depends on the size of the QuIDD representation of
%% $p(\cdot)$, which is small in many important cases. For example,
%% the original paper by Grover \cite{Grover97} assumes that a unique
%% solution to $p(x)=1$ exists. In that case, the QuIDD of $p(\cdot)$
%% has only $k$ nodes, and QuIDD-based simulation is very fast.
%%  This result demonstrates exploitable structure
%% in some supposedly unstructured black-boxes. In fact, Grover's
%% algorithm can be simulated efficiently for any $p(\cdot)$ captured
%% by a QuIDD with polynomially many nodes.

  Predicates exist that do not compress well in QuIDD form,
  and so require super-polynomial time and memory resources.
  However, some of these predicates may also require
  a super-polynomial number of quantum gates. This may cause the
  evaluation of $p(x)$ to dominate the runtime of quantum search and
  undermine the speed-up of Grover's algorithm over classical search.

%%   Fast classical simulation is another competitor of quantum
%%   algorithms. The QuIDDPro simulator discussed in Section
%%   \ref{sec:fair} can automatically detect structure in input states
%%   and quantum circuits, and then take advantage of it by means of data
%%   compression.  Indeed, it is hard to imagine a useful search
%%   application with absolutely no structure.  If $p(\cdot)$ in Grover's
%%   algorithm implements a randomly-chosen predicate, then the smallest
%%   quantum and classical circuits for it are guaranteed to be
%%   exponential {because there are many more functions than small
%%   circuits}. Such a large circuit may dwarf the Grover speed-up.
%%   In another case, when a non-trivial fraction of all
%%   inputs satisfy $p(\cdot)$, classical randomized search is very fast.
%%   If $p(\cdot)$ allows only one satisfying solution (or a limited
%%   number of them), then its QuIDD must be very small and QuIDD-based
%%   simulation will run nearly as fast as a quantum computer. To this
%%   end, Valiant and Vazirani proved \cite{VV} that NP-complete problems
%%   can be reduced to those with unique solutions.  An important caveat
%%   is that constructing a small QuIDD from an alternative description
%%   of $p(\cdot)$ may require significant time and space, e.g., if we
%%   search for integers $p, q$ such that $pq=n$ for a
%%   difficult-to-factor $n$.

\vspace{-2mm}
\section{Problem-Specific Algorithms}
\vspace{-2mm} \label{sec:showdown}

% A theoretical improvement due to quantum computation may be just a proof
% of concept (cf. Deutsche's algorithm \cite{Deutsch,NielsenC})
% and does not necessarily lead to observable speed-up in useful applications.
%   As evidenced by the discussion of the first two requirements,

  As discussed above, comparisons between quantum and classical search
algorithms often implicitly make strong assumptions, namely { the
 unrestricted use of black-box predicates} and the misconception that
{ no quantum circuit can be simulated efficiently on any inputs}.
{\em These assumptions overestimate the
potential speed-up offered by quantum search}. Another, and perhaps, more
serious oversight in popular analysis concerns the structure present in
particular search problems.  We show next that one must compare
Grover's algorithm against highly-tuned classical algorithms specialized
to a given search problem, rather than against generic exhaustive search.

% that have in the past been suggested as applications of quantum search.

     Boolean 3-satisfiability and graph 3-coloring
  have been suggested as possible applications of quantum search because
  polynomial-time algorithms are not known for these NP-complete
  problems, and are unlikely to be found. There are, however,
  classical algorithms \cite{Schoning99,Eppstein01} which solve
  these two problems in less than $\sim poly(k) 1.34^k$ and $\sim
  poly(k) 1.37^k$ steps, respectively ($k$ is the number of
  variables), thereby outperforming quantum techniques that require
  at least $\sim poly(k) 1.41^k$ steps.\footnote{The algorithm
  analyzed in \cite{Schoning99} is a simplified version
  of the well-known WalkSAT program. It is a type of randomized local search
  where variable assignments are changed one at a time so as to
  statistically decrease the number of unsatisfied clauses.
  Randomization in the algorithm facilitates hill-climbing
  and an extremely fast move selection mechanism. }
  The algorithms in \cite{Schoning99,Eppstein01} exploit subtle structure in problem
  formulations, and so have the potential for further improvement.
  No such improvement is possible for Grover's algorithm
  unless additional assumptions are made \cite{Zalka99},
  as in \cite{RolandC03}.

  Of course, many practical NP-complete search problems remain
  whose best known upper bounds far exceed $\sim poly(k)1.41^k$ --- for
  example, $k$-satisfiability and $k$-coloring, with $k\geq 4$.
  However, known classical algorithms often finish much faster
  on certain inputs, both
  application-derived and artificial, structured and unstructured
  \cite{Kautz03}. Indeed, one can now solve randomly-generated
  hard-to-satisfy instances of Boolean satisfiability with a million
  variables in one day, using a single-processor PC.
  Grover's algorithm is mainly sensitive to the number of solutions,
  but not to the solutions themselves and not to input features (such
  as symmetries) that are sometimes exploited by classical algorithms.

    The Euclidean traveling salesman problem (TSP) and many other
  geometric optimization problems have defied fast exact algorithms so far,
  but can often be solved using polynomial-time approximation schemes
  that trade off accuracy for runtime \cite{Arora96}.
  Their geometric structure allows one to solve these problems
  to a given precision $\epsilon>0$ in polynomial time. Additionally,
  specific very large instances of such problems have been solved optimally
  in the past, e.g., the TSP for over 10,000 cities.
  Opportunistic algorithms and heuristics that work well only on some inputs
  are very useful in practice, but comparable quantum heuristics are poorly
  understood.

  Hard cryptography problems have also been mentioned as potential
  applications of Grover's algorithm \cite{Zalka00}. They include
  code-breaking (particularly, the DES and AES cryptosystems) and
  reversing cryptographically-secure hash-functions (MD5 and SHA-1).
  Indeed, these can be cast as unstructured search algorithms, but
  cryptographers have identified  structure in all such
  applications. The task of breaking DES has been reduced to Boolean
  3-satisfiability \cite{MarraroM99} and clearly does not require a
  naive enumeration of keys.  ``Essential algebraic structure''
  \cite{MurphyR02} has been identified in AES, which has recently
  replaced DES as the US encryption standard.  Similarly, publications
  surveyed in \cite{Robshaw02} suggest that all major crypto-hashes
  have well-pronounced structure. In fact, past discoveries of unexpected
  types of structure in crypto-hashes MD-2, MD-4, RIPEMD and SHA were
  so significant that those functions are no longer considered
  cryptographically secure.

  In summary, to evaluate the potential benefits of an implementation of
  Grover's algorithm, one must compare it with the best known classical
  problem-specific algorithms taking exploitable structure into account.
  The comparison is straightforward if the
  search problem has been studied previously, as in the case of
  Boolean satisfiability, graph coloring, the traveling salesman problem,
  and various code-breaking tasks.
  However, even if no optimized classical algorithm
  has yet been devised for a particular search problem, the problem
  may still contain a great deal of implicit and exploitable structure.

%\vspace{-3mm}
\section{Conclusions}
\label{sec:conclusions}
% \vspace{-3mm}

% {\em We take it for granted that quantum computers
%   may be immensely useful in physically-secure quantum communication
%   and numerical simulation of quantum mechanics, e.g., in astrophysics
%   and nuclear engineering.}

  While quantum computing has dramatically advanced through the last decade,
  its potential applications have not yet been demonstrated at full scale.
  Such demonstrations are likely to require breakthroughs in physics,
  computer science and engineering \cite{Preskill98}.
  Additionally, it is important to understand current roadblocks
  to achieving practical speed-ups with quantum algorithms.
  To this end, we have analyzed the potential of Grover's search algorithm
  to compete with classical methods for search, and identified three requirements
  for it to be practically useful. They serve to highlight several
  specific obstacles and pitfalls in implementing and analyzing quantum search,
  e.g., ignoring the implementation complexity of the query process.
  We have also demonstrated the usefulness of classical simulation
  in evaluating quantum algorithms and their implementation.
  Our hope is that this work will temper unreasonable expectations of quantum
  speed-ups and encourage further study of ways to improve quantum search.

  In the near term, quantum computers running Grover's algorithm are
  unlikely to be competitive with the best classical computers in
  practical applications. Adding to the arguments of \cite{Zalka00}
  regarding classical parallelism, we have pointed out that
  recent work on solving ``intractable'' problems such as Boolean
  satisfiability and direct simulation of quantum circuits, offers
  opportunities to exploit subtle domain-specific structures, even on a
  single classical processor. Despite their exponential worst-case runtime,
  some of these algorithms are always faster than Grover's search.
 
  We list some interesting open research issues that deserve attention:

  \begin{enumerate}
\vspace{-1mm}
    {\item Applications of search where classical methods do not offer
   sufficient scalability.}
\vspace{-1mm}
    {\item Algorithms for near-optimal synthesis of quantum oracle circuits.}
\vspace{-1mm}
    {\item Quantum heuristics that finish faster or produce better solutions on practical inputs.}
\vspace{-1mm}
    {\item Quantum algorithms that exploit the structure of useful search
problems.}
\vspace{-1mm}
  \end{enumerate}

   Recent work on variants of Grover's search that account
   for problem structure seems particularly promising.
   Building up on earlier proof-of-concept results,
   Roland and Cerf \cite{RolandC03a} compare
   the fastest known classical algorithms for 3-satisfiability
   \cite{Schoning99} to their quantum search algorithm cognizant
   of the 3-literal limitation. Their analysis shows that the quantum
   algorithm has a smaller asymptotic expected runtime, averaged
   over multiple SAT instances with a particular clause-to-variable ratio
   (at the phase-transition), which are known to be the most difficult 
   to solve on average.  Similar comparisons for worst-case asymptotic runtime
   remain an attractive goal for future research.

   Several considerations in our work are not restricted to Grover's search
   and apply to other potential  applications of quantum computing.
   For example, the {\em graph automorphism} problem that seeks symmetries
   of a given graph is sometimes suggested as a candidate for polynomial-time
   quantum algorithms \cite{Preskill98}. Classically this problem
   appears to require more than polynomial time in the worst case, while
   unlikely to be NP complete, just like number-factoring. However,
   graph automorphism is provably easy for random graphs and can also
   be solved quickly in many practical cases, e.g., in the context of microprocessor
   verification \cite{DargaSM04}, making existing algorithms and software strong
   competitors of potential future quantum algorithms.

   A novel application of quantum algorithms to finding template matches
   in photographs was proposed in \cite{CurtisM03}, but relies on the quantum Fourier
   transform rather than Grover's search. Proposals of this kind also deserve careful
   evaluation from the computer engineering perspective and must be compared
   to state-of-the-art classical methods on realistic inputs.

   {\bf Acknowledgments.} This work is supported in part by DARPA and NSF.
   The views and conclusions contained herein are those of the authors
   and should not be interpreted as necessarily representing official
   policies or endorsements of their employers or funding agencies.

%% {\em The reference list needs to be pruned and the style, e.g.
%% capitalization,  made uniform.}

%\vspace{-2mm}
  {\bf Note.} In the list of references below, {\tt quant-ph} refers to
  quantum physics abstracts available online from \\
   {\tt http://arxiv.org/abs/quant-ph}

\section*{Appendix: Description of Grover's Quantum Search Algorithm}

Grover's quantum algorithm searches for a subset of items in an
unstructured set of $N$ items \cite{Grover97}.  The algorithm
incorporates the search criteria in the form of a black-box
predicate that can be evaluated on any items in the set. The
complexity of this evaluation (query) varies depending on the
search criteria. With conventional algorithms, searching an
unstructured set of $N$ items requires $\Omega(N)$ queries in the
worst case.
% If the number of items matching the search criteria is
% known a priori, the complexity of unstructured search becomes
% $O(N)$.
 In the quantum domain, however, Grover's algorithm can
perform unstructured search by making only $O(\sqrt{N})$ queries,
a quadratic speed-up over the classical case.  This improvement is
contingent on the assumption that the search predicate can be
evaluated on a superposition of all database items. Additionally,
converting classical search criteria to quantum circuits often
entails a moderate overhead, and the complexity of the quantum
predicate can offset the reduction in the number of queries.

\begin{figure*}[tb]
\begin{center}
\includegraphics[width=16cm]{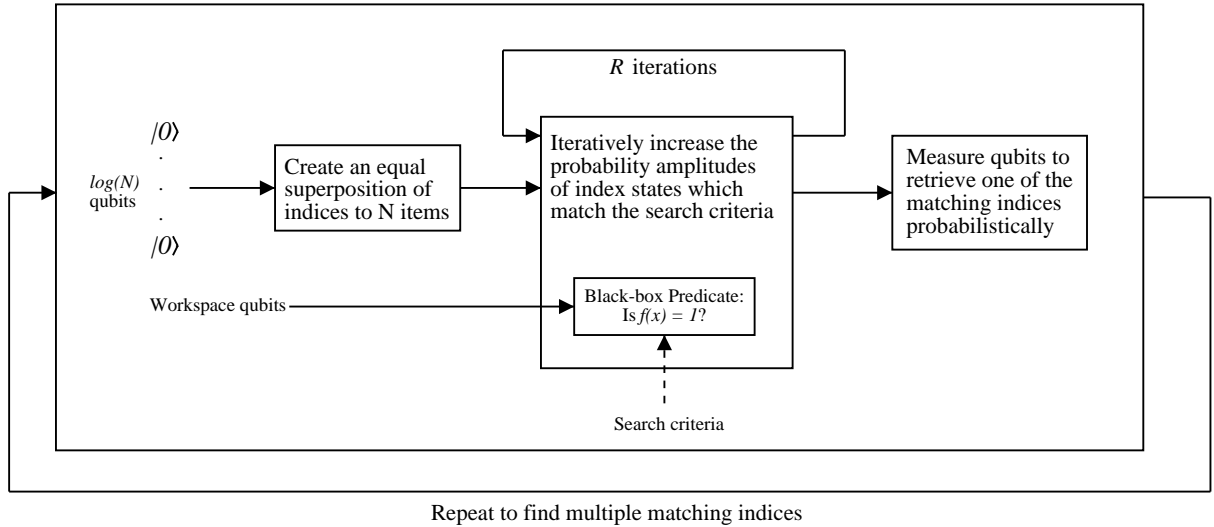}
%\vspace{-5mm}
\parbox{14cm}{\caption{\label{fig:grover_high_level} %\small
   A high-level circuit depiction of Grover's quantum search algorithm. More detailed circuits can be found in \cite{NielsenC} and \cite{ViamontesMH03}.}}
%\vspace{-2mm}
\end{center}
\end{figure*}

A high-level circuit representation of Grover's algorithm is shown
in Figure \ref{fig:grover_high_level}. The first step of the
algorithm is to initialize $log(N)$ qubits, each with a value of
$|0 \rangle$. These qubits are then placed into an equal
superposition\footnote{An equal superposition means that all
possible states represented by the superposition have equal
probability amplitudes. The square of the probability amplitude
associated with a particular bit-string is the probability with
which that bit-string will be observed after a quantum
measurement.} of all values from $0$ to $N - 1$ (encoded in
binary), by applying one Hadamard gate on each input qubit. Since
the superposition contains bit-strings, they are thought of as
indices to the $N$ items in the search space rather than as the
items themselves.

The next step is to iteratively increase the probability
amplitudes of those indices in the superposition that match the
search criteria. The key component of each iteration is querying
the black-box predicate. The predicate can be viewed abstractly as
a function $f(x)$ which returns $1$ if the index $x$ matches the
search criteria and returns $0$ otherwise. Assuming that the
predicate can be evaluated on the superposition of indices, a
single query then evaluates the predicate on all indices
simultaneously. In conjunction with extra gates and qubits
(``workspace qubits''), the indices for which $f(x)=1$ can be
marked by rotating their phases by $\pi$ radians. To capitalize on
this distinction, additional gates are applied so as to increase
the probability amplitudes of marked indices and decrease the
probability amplitudes of unmarked indices. Mathematically, this
transformation is a form of inversion about the mean. It can be
illustrated on sample input \{-1/2, 1/2, 1/2, 1/2\} with mean 1/4,
where inversion about mean produces \{1,0,0,0\}. Note that both
vectors in the example have norm 1, and in general inversion about
mean is a unitary transformation. One of Grover's insights was
that it can be implemented with fairly small quantum circuits.

In the case when only one element out of $N$ satisfies the search
criterion each iteration of Grover's algorithm increases the
amplitude of this state by approximately $O(1/\sqrt{N})$. Therefore,
on the order of $\sqrt{N}$ iterations are required to maximize the
probability that a quantum measurement will yield the sought index
(bit-string). For the more general case with $M$ elements
satisfying the search criteria, the optimal number of iterations
is shown in \cite{Boyer96} to be $R = \frac{\pi \
arcsin(\sqrt{M/N})}{4}$. It can also be shown that Grover's
algorithm exhibits a periodic behavior and after the amplitudes of
sought elements peak, they start decreasing.

The final step is to apply a quantum measurement to each of the
$log(N)$ qubits. Postulates of quantum mechanics posit that
measurement is probabilistic and collapses the superposition to a
single bit-string --- the larger the amplitude of a bit-string,
the more likely the bit-string is to be observed.

 When $M$ items match the search criteria in a particular search problem,
 then Grover's algorithm produces one of them. Each item is equally likely
 to appear because the inversion about the mean process increases
 the probability amplitudes of matching items equally.
 If all such items must be found, Grover's algorithm may have to be
 repeated more than $M$ times, potentially returning some items more
 than once. On the other hand, classical deterministic search
 techniques avoid such duplication and may be more suitable in
 applications where $M$ is a significant fraction of $N$.

\end{document}